\documentclass[fleqn,usenatbib]{mnras}

\usepackage{newtxtext,newtxmath}

\usepackage[T1]{fontenc}
\usepackage{ae,aecompl}

\usepackage{graphicx}	
\usepackage{amsmath}	

\def\ione{\,{\sc i}}
\def\ii{\,{\sc ii}}

\def\vmicro{$\xi_{\rm t}$}

\newcommand{\kms}{km\,s$^{-1}$}
\newcommand{\te}{$T_{\rm eff}$}
\newcommand{\logg}{$\log{g}$}
\newcommand{\vsini}{$v\sin{i}$}
\def\vmacro{$\zeta_{\rm RT}$}
\def\Msun{$M_{\odot}$}

\def\Rsun{$R_{\odot}$}
\def\CT{$T_{\rm c}$}
\title[Abundance analysis of the SB2 system HD~60803]
{Fundamental parameters and abundance analysis of the components in the SB2 system HD~60803}

\author[T. Ryabchikova et al.]{
\parbox{\textwidth}{
T.~Ryabchikova,$^1$\thanks{E-mail: ryabchik@inasan.ru} S.~Zvyagintsev,$^{1}$ A.~Tkachenko,$^{2}$ V.~Tsymbal,$^{1}$ Yu.~Pakhomov,$^{1}$ E.~Semenko$^{3,4}$}\\\\
$^{1}$Institute of Astronomy of the Russian Academy of Sciences, Pyatnitskaya str. 48, 119017, Moscow, Russia \\
$^{2}$Institute of Astronomy, KU Leuven, Celestijnenlaan 200D, 3001, Leuven, Belgium \\
$^{3}$Special Astrophysical Observatory, Russian Academy of Sciences, 369167 Nizhny Arkhyz, Russia, \\
$^{4}$National Astronomical Research Institute of Thailand, 260 Moo 4, T. Donkaew, A. Maerim, 50180 Chiangmai, Thailand
}

\date{Accepted XXX. Received YYY; in original form ZZZ}
\pubyear{2021}

\begin{document}
\label{firstpage}
\pagerange{\pageref{firstpage}--\pageref{lastpage}}
\maketitle

\begin{abstract}
We performed a detailed spectroscopic study of the SB2 system HD\,60803 based on high-resolution spectra obtained with the different spectrographs. The analysis was done with two independent methods:  a) the direct modelling of the observed binary spectrum by a sum of synthetic spectra varying a set of free parameters and minimizing a difference between the observed and theoretical spectra; b) spectrum disentangling and an independent  modelling of the individual components. Being applied to binary spectra from different spectrographs both methods converge to a consistent solution for the fundamental parameters of the HD~60803 components: \te=6\,055$\pm$70~K,  \logg=4.08$\pm$0.12, \vmicro=1.45$\pm$0.18~\kms, [M/H]=0.03$\pm$0.06 (primary), and \te=6\,069$\pm$70~K, \logg=4.14$\pm$0.09, \vmicro=1.48$\pm$0.18~\kms, [M/H]=0.03$\pm$0.06 (secondary). Differential abundance analysis of the components did not reveal any significant difference in their chemical composition. Besides Li both components have solar atmospheric abundances.  Li abundance exceeds the solar one by $\sim$2~dex, but it agrees with Li abundance in main-sequence late F-stars. Relative-to-solar abundances in both components slightly correlate with the condensation temperature the same way as was found in the solar analogs with/without detected giant planets. The estimated age of the system is 5.5$\pm0.5$~Gyr.    
\end{abstract}




\section{Introduction}

The study of atmospheric abundances of the solar-like binary system components has a special interest in the light of the planet systems formation. The discovery of the planets around one of the component of the wide binary system gave start to differential abundance analyses of the wide binaries with the aim to discover 'fingerprints' of the confirmed or possible planets and planet formation. However, the results were controversial. For one of the best studied wide binary 16~Cyg with the giant gas planet orbiting around 16~Cyg~B \citep{1997ApJ...483..457C} no abundance difference between the components were derived by \citet{2011ApJ...737L..32S}, while a constant difference of +0.04~dex (A-B) for all elements was found by \citet{2011ApJ...740...76R} without trend with the condensation temperature \CT. The overall abundance difference as well as the correlation with \CT\ are usually attributed to the chemical signature of planet formation.  
The result of \citet{2011ApJ...740...76R} that 16~Cyg~A is more metal rich than 16~Cyg~A was confirmed later on the basis of  better quality observations obtained with two different spectrographs \citep{2014ApJ...790L..25T, 2019A&A...628A.126M}.
However, the authors got different trends in relative abundances (A-B) as a function of \CT, that leads to different interpretation of the planet formation.  \citet{2001A&A...377..123G} performed differential abundance analysis of six wide pairs (separations of few hundreds AU) without detected planets. They did not find abundance difference between components in four systems and detect significant difference in two remaining binary systems HD~219542 and HD~200466. 
Later, \citet{2004A&A...420..683D,2006A&A...454..581D} extended differential analysis to 56 wide pairs with 6 systems in common. Their analysis was restricted to atmospheric parameters and differential iron abundance determinations. HD~219542 was included in both studies and the authors did not find iron abundance difference between the components.
 
Higher metallicity of one of the components in wide binaries is often explained as the accretion of rocky planets or the inner dust-rich part of the protoplanetary disk onto the star \citep[see, for example,][]{2001A&A...377..123G}. Dynamical calculations by \citet{2000ApJ...545.1064L} show that the stars in short-period binaries (spectroscopic SB2) may protect themselves from the metallicity enhancement caused by accretion process. If it's a case then the metal-rich stars in SB2 systems are rare events compared to single planet-hosting stars or those in wide binaries. Hence, it is important to extend the differential analysis to other binary systems with different separations.

HD\,60803 (HR~2918) is a bright system discovered to have double-lined spectrum by \citet{1964ApJ...140.1401W}. The orbit solution was obtained by \citet{1997Obs...117..208G}. The system consists of almost equal pair of stars with the masses slightly exceeded the solar mass. The orbital period of the system is $P$=26.1889 days. \citet{2014AJ....147...86T} gives $M_1$=1.31\Msun\ and $M_2$=1.28\Msun\, in his catalogue of binary stars. Effective temperature, surface gravity and metallicity of the system were obtained by different methods of photometry and spectroscopy. \citet{2006A&A...450..735M} derived \te=5934(53)~K from 2MASS photometry. Similar values, \te=5933~K and \te=5945~K were derived from IR flux modelling  \citep{2012MNRAS.427..343M} and from spectroscopic analysis \citep{2013A&A...553A..95M}. The latter used an automatic PCA (principal component analysis) procedure applied to {\sc ELODIE} spectrum (see next Section~\ref{sect:obs}). Another automatic spectroscopic analysis of the {\sc ELODIE} spectra  which uses SP\_Ace (Stellar Parameters And Chemical abundances Estimator) code \citep{2016A&A...587A...2B} provided slightly lower effective temperature and rather low metallicity. Independent analysis of two {\sc ELODIE} spectra resulted in the following parameters: \te=5853~K, \logg=3.52, [M/H]=--0.64;  \te=5913~K, \logg=3.65, [M/H]=--0.63, respectively. The latest version of the  Geneva-Copenhagen survey contains atmospheric parameters (\te=6002~K, \logg=3.87, [Fe/H]=+0.11) as well as the  age estimate $t$=3.2~Gyr \citep{2011A&A...530A.138C}. Higher age $t$=3.93~Gyr is reported in \citet{2010ApJ...725..875I}. 

All the above mentioned parameter estimates are based on the study of the HD\,60803 SB2 system  as a single star without separation into individual components. Obviously, both stars have been formed from the same cloud, therefore one can expect that their atmospheric abundances are identical. The separation between the components is $\sim$25\Rsun\, that allows to consider the components evolution as an evolution of single stars. Therefore by splitting the binary spectrum into individual spectra of the components we can analyse them as spectra of single stars and to perform a differential abundance determinations. 

The paper is organized as follows. Observations are described in Section~\ref{sect:obs}.  The results of atmospheric parameters and abundance determinations based on the analysis of binary spectrum at one orbital phase are given in Section\ref{sect:single}. Section~\ref{sect:SLD} provides a short description of disentangling process applied to a set of 5 binary spectra. The results  of atmospheric parameters and abundance determinations performed for individual spectra of both components are presented in Section~\ref{sect:abn}. Masses, age and orbital inclination of the binary system HD\,60803 are given in Section~\ref{sect:evol}.     

\section{Observations}\label{sect:obs}

Two spectra of HD\,60803 taken at the phases of maximum radial velocity (RV) separation of the individual binary components were extracted from the archives \footnote{http://atlas.obs-hp.fr/elodie/}$^,$\footnote{https://koa.ipac.caltech.edu/cgi-bin/KOA/nph-KOAlogin}. One of the spectra was obtained with the {\sc HiRes} spectrograph attached to the 10-m Keck telescope while the second spectrum was taken with the  {\sc ELODIE} spectrograph mounted on the 1.93-m telescope of the Observatoire de Haute-Provence. The spectrum taken with the {\sc HiRes} instrument covers three wavelength ranges between 3\,500 and 4\,730~\AA\ (blue), 4\,900 and 6\,400~\AA\ (green), and 6\,540 and 7\,990~\AA (red). The {\sc ELODIE} spectrum provides the wavelength coverage between 4\,000 and 6\,800~\AA. The resolving power achieved is 42\,000 and 69\,000 in the case of the {\sc ELODIE} and {\sc HiRes} spectrographs, respectively. In both cases the data were reduced using dedicated pipelines and normalized to the local continuum afterwards.

Four spectra of HD\,60803 were obtained in March 2017 with an echelle spectrograph {\sc NES} \citep{2017ARep...61..820P} mounted at the Nasmyth focus of the Russian 6-m telescope BTA of the Special Astrophysical Observatory. Information about the time of observation and spectral resolution of extracted spectra are available in the Table~\ref{tab1}. A single set of observational data includes one spectrum of the star, twenty frames for bias subtraction and flat field correction, and one spectrum of a ThAr lamp for wavelength calibration. Raw CCD images were reduced with the help of the package \textsc{echelle} of ESO-MIDAS tools \citep{1992ASPC...25..115W}. 50 spectral orders were extracted in the range of 4047-7081~\AA. Then they were normalized to the continuum level and merged to one 1D spectrum. An averaged signal-to-noise (S/N) ratio is $\approx$250. 

Five spectra of HD\,60803 were obtained in October 2018 with the {\sc HERMES} spectrograph \citep{2011A&A...526A..69R}, mounted on the 1.2-m Mercator telescope on La Palma, Canary Islands, Spain. The {\sc HERMES} spectra cover a wavelength range between 3750 and 9000~\AA\ with a spectral resolving power of $R$=85\,000. The wavelength calibration was obtained from emission spectra of thorium-argon-neon lamp. Data reduction of the observed stellar spectra were performed with the instrument-specific pipeline \citep{2011A&A...526A..69R}. Details of the {\sc HERMES} observations are presented in Table~\ref{tab1}. 

\begin{table}
\footnotesize
\centering
\tabcolsep=1.0mm
\caption{Observational journal of HD~60803. MJDs, orbital phases, primary's and secondary's RV measurements are given. Information on spectrographs is provided in the last two columns.}\label{tab1}
\begin{tabular}{ccllcc}\hline
MJD       &Orbital&\multicolumn{2}{c}{RV, \kms}& Spectrograph & $R$ \\
2400000+  &phase  & Primary & Secondary &      &                  \\
\hline
50827.9898  & 0.176 &-42.96(09)	&~53.08(09) & ELODIE & 42\,000 \\
53696.5400	& 0.709	&~46.52(09) &-38.81(09) & HiRes  & 69\,000 \\
57819.6659  & 0.147 &-44.45(19)	&~55.94(19) & NES    & 57\,000 \\
57820.6606  & 0.185 &-39.03(19)	&~52.73(19) & NES    & 57\,000 \\
57821.6545  & 0.223 &-35.13(19)	&~44.03(19) & NES    & 57\,000 \\
57821.8625  & 0.231 &-33.50(19)	&~42.56(19) & NES    & 57\,000 \\
58404.1805	& 0.466	&~14.48(06) &~-4.98(05) & HERMES & 85\,000 \\
58405.1552	& 0.503	&~22.55(06) &-12.00(05) & HERMES & 85\,000 \\
58406.1524	& 0.541	&~28.53(06) &-18.68(05) & HERMES & 85\,000 \\
58407.1613	& 0.580	&~34.89(06) &-25.00(05) & HERMES & 85\,000 \\
58409.1878	& 0.657	&~44.01(06) &-34.25(05) & HERMES & 85\,000 \\
\hline                   
\end{tabular}
\end{table}


\section{Spectral analysis }
There are two approaches to the spectral analysis of SB2 systems. The first one considers the direct modelling of the observed binary spectrum by a sum of synthetic spectra varying a set of free parameters and minimizing a difference between the observed and theoretical spectra. In the second method the observed spectra obtained at different orbital phases are disentangled in the individual spectra of the components. 
    
We applied both approaches in our work. In the beginning we have only two spectra at our disposal obtained with different spectrographs {\sc ELODIE} and {\sc HiRes} at approximately opposite orbital phases. It was not possible to disentangle these spectra therefore they were used mainly for radial velocities (RV) measurements and atmospheric parameters estimates. Later we get two sets of observations with {\sc NES} and {\sc HERMES} spectrographs. The first set was not suitable for spectrum disentangling due to small phase range, and it was used for RV measurements only. Spectrum disentangling procedure was successfully applied to {\sc HERMES} spectra that allowed us to perform an accurate abundance analysis of both components of the HD\,60803 binary system.

\subsection{ {\sc ELODIE} and {\sc HiRes} spectra}\label{sect:single}


We used the methods of model atmospheres and spectrum synthesis as implemented in the Grid Search in Stellar Parameters \citep[GSSP;][]{2015A&A...581A.129T} software package to perform detailed analysis of both spectra of HD\,60803. The atmosphere models were pre-computed with the {\sc LLmodels} code \citep{2004AA...428..993S}; details on the available grid of models can be found in \citep{2012MNRAS.422.2960T}. We use the {\sc SynthV} radiative transfer code \citep{1996ASPC..108..198T} to synthesize the spectra in an arbitrary wavelength range. Both the atmosphere model and the radiative transfer codes are based on the assumption of local thermodynamical equilibrium (LTE) which is justified in the case of the HD\,60803 system. The information about atomic data for the lines of interest is extracted from the  Vienna Atomic Line Database ({\sc VALD}) \citep{1999AAS..138..119K}.

\begin{figure}
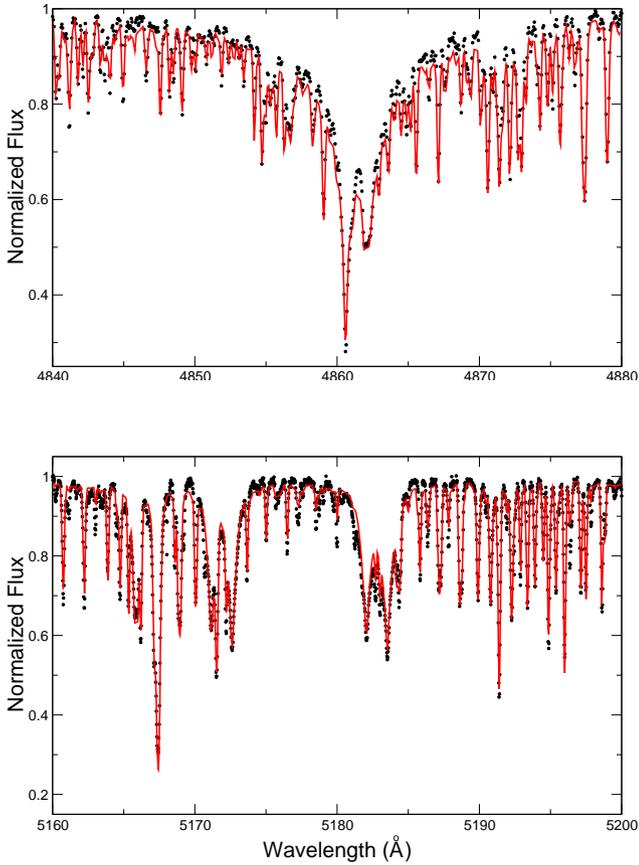

	\includegraphics[scale=0.32]{elodie.eps}\vspace{5mm}
	\includegraphics[scale=0.32]{hires.eps}
	\caption{Quality of the fit to the composite spectrum of the
		HD\,60803 system. The observed spectrum is indicated by black
		dots, while red line refers to the best fit synthetic
		spectrum. Top:  {\sc ELODIE} spectrum. Bottom: {\sc HiRes} spectrum.}
	\label{E+HR}
\end{figure}

The {\sc gssp} software package is based on the grid search in different atmospheric parameters, such as the effective temperature (\te), surface gravity (\logg), micro- (\vmicro) and macro-turbulent (radial-tangential \vmacro) velocities, projected rotational velocity (\vsini), and global metallicity ([M/H]) of the star. Optimization of the individual abundances of different chemical elements is also possible. The {\sc gssp} package is a compilation of three independent modules, each one focusing on particular analysis tasks. The {\sc gssp\_single} module is designed for the analysis of spectra of single stars and is also applicable to the disentangled spectra of individual binary components. In the latter case, the code takes care of possible light dilution of the spectra which however is assumed to be wavelength independent. On contrary, the {\sc gssp\_binary} software module is specifically designed for simultaneous analysis of light diluted disentangled spectra of both binary components, where wavelength dependence of the light dilution effect is taken into account. In this particular case, the ratio of the radii of two stellar components is optimized instead of the wavelength-independent dilution factor as is implemented in the {\sc gssp\_single} module. The directly observed composite spectra of double-lined spectroscopic binary systems can be analysed with the {\sc gssp\_composite} module. The methodology is similar to the one implemented in the {\sc gssp\_binary} module, where the light dilution is assumed to be wavelength-dependent and is taken into account by optimizing the ratio of the radii of the two stellar components. In addition, the individual RVs can be set as free parameters. Without having at our disposal the disentangled spectra of the individual stellar components of the HD\,60803 system, the analysis was restricted to the observed composite spectra and was performed by means of the {\sc gssp\_composite} module.

{\sc gssp\_composite} starts from computing individual grids of synthetic spectra per binary component. A possible overlap in the two grids is taken into account to avoid repetitive calculations of synthetic spectra. All possible combinations are then built between the two grids of synthetic spectra, resulting in theoretical composite spectra which are then compared to the observations. That is, the grids of respectively 10 and 15 synthetic spectra for the primary and secondary components of a binary system result in a total of 150 possible combinations that will be finally processed. We use the $\chi^2$-merit function to judge on the goodness of fit; the error bars are 1-$\sigma$ uncertainties computed from $\chi^2$ statistics by taking into account possible correlations between different atmospheric parameters. More information on the {\sc gssp} software package and error calculations can be found in \citet{2015A&A...581A.129T}.

The Least Squares Deconvolution \citep[LSD;][]{1997MNRAS.291..658D} method was used to compute average profiles from both observed spectra of the HD\,60803 system. The method assumes that all lines in the spectrum of a given star have similar shape and that they add-up linearly. For this reason, the lines with developed wings (e.g., Balmer lines, some of the magnesium lines in the case of solar-type stars, etc.) are excluded from the calculations and the LSD profile represents an average profile of all the remaining metal lines in the considered wavelength interval. This way, the LSD method is similar to the method of discrete cross-correlation, where the observed spectrum is cross-correlated with a set of delta functions. In our case, a list of atomic lines (``line mask'') is extracted from the  {\sc VALD} database assuming a particular set of atmospheric parameters. The computed average profiles are used by us for the determination of precise RVs and to study the details of line broadening mechanisms for both stellar components of the HD\,60803 system.

Initially, the LSD profiles were extracted from he observations using the line mask computed under the assumption of solar chemical composition and effective temperature and surface gravity of 5\,900~K and 4.0~dex, respectively, for both binary components. These LSD profiles gave a good estimate for the RVs of both binary components which were computed as the center of gravity (first order moment) of the corresponding profile. Fixing the RVs of both stars during the spectrum analysis with the {\sc gssp\_composite} module is advantageous as this significantly lowers the total number of synthetic spectra combinations that have to be processed during a single run. The following atmospheric parameters were set as free parameters for each of the binary components during the analysis of the {\sc ELODIE} spectrum with the {\sc gssp\_composite} software module: \te, \logg, \vsini, \vmicro, [M/H], and the radii ratio (R$_A$/R$_B$). Macro-turbulent velocity was fixed to 3~\kms\ for both components, the average value computed from the individual measurements obtained from the two different spectra,  {\sc ELODIE} and  {\sc HiRes}. The analysis was restricted to the wavelength range between 4\,500 and 6\,000~\AA, as the continuum normalization becomes more uncertain towards the blue part of the spectrum, whereas in its red part the spectrum gets strongly blended with telluric contributions. The results of the spectrum analysis are summarized in Table~\ref{tab_param}; a small portion of the observed spectrum and the best fit model are illustrated in Figure~\ref{E+HR} (top panel).

The  {\sc HiRes} spectrum does not provide a wavelength coverage around any of the Balmer lines except H$_{\alpha}$. Moreover, H$\alpha$ profile is placed right at the edge of the first echelle order in the red part of the spectrum, which makes the normalization of this profile very uncertain. For these reasons, the analysis of the  {\sc HiRes} spectrum was restricted to metal lines only; similar to the case of the  {\sc ELODIE} spectrum, we stick to the total wavelength coverage between 4\,500 and 6\,000~\AA\ and optimize all size atmospheric parameters: \te, \logg, \vsini, \vmicro, [M/H], and R$_1$/R$_2$, with macroturbulence fixed to 3~\kms. The results of the analysis are summarized in Table~\ref{tab_param}; the quality of the fit in a selected wavelength region is illustrated in Figure~\ref{E+HR} (bottom panel).

Finally, we optimized \vsini\ and \vmacro\ values. The fundamental parameters extracted from the  {\sc ELODIE} spectrum were used to compute line masks for both binary components in order to extract LSD profiles from the two available observed composite spectra. Grids of synthetic spectra were computed for the primary and secondary components of the HD\,60803 system by fixing \te, \logg, and [M/H] to the values extracted from the  {\sc ELODIE} spectrum, but varying \vsini\ and \vmacro\ values. Synthetic LSD profiles were extracted from these grids of synthetic spectra assuming the same wavelength range that was used to compute LSD profiles from the observations. We then used the ratio of the radii of the two components reported in Table~\ref{tab_param} (the {\sc ELODIE} results) to compute composite synthetic LSD profiles from all possible combinations of the theoretical spectra from the individual grids. The obtained theoretical composite LSD profiles were compared to the observed ones by varying RVs of the individual binary components in addition to varying their \vsini\ and \vmacro\ values. We used $\chi^2$ merit function to judge on the goodness of the fit; 1-$\sigma$ error bars were computed from $\chi^2$ statistics. The results are summarized in Table~\ref{tab_param}.

\begin{table}
	\centering\tabcolsep 1.5mm\caption{Atmospheric parameters of both stellar components of the HD\,60803 system obtained from fitting the observed composite spectra taken with three different instruments. Error bars given in parentheses represent 1$\sigma$ uncertainties.}\label{tab_param}
\begin{tabular}{llcc} \hline
	\multicolumn{1}{c}{Parameter\rule{0pt}{9pt}} &
	\multicolumn{1}{c}{Unit} & \multicolumn{1}{c}{Primary(A)} &
	\multicolumn{1}{c}{Secondary(B)}\\\hline
\multicolumn{4}{c}{{\bf  {\sc ELODIE} spectrum ({\sc gssp\_composite})\rule{0pt}{11pt}}}\\
\te\rule{0pt}{15pt} & K    & 5\,750(170) & 5\,850(200)\\
\logg               & dex  & 3.75(0.30)  & 3.92(0.35)\\
\vmicro              & \kms & 1.35(0.60)  & 1.10(0.60)\\
\vsini              & \kms & 1.0(2.0)    & 1.0(1.9)\\
\vmacro             & \kms & 3.3(1.2)    & 3.5(1.2)\\
${\rm [M/H]}$       & dex  & $-$0.20(0.10)& $-$0.10(0.10)\\
R$_A$/R$_B$         &      & \multicolumn{2}{c}{1.09(0.10)}\\
\multicolumn{4}{c}{{\bf  {\sc HiRes} spectrum ({\sc gssp\_composite})\rule{0pt}{11pt}}}\\
\te\rule{0pt}{15pt} & K    & 6\,050(180) & 6\,090(190)\\
\logg               & dex  & 4.20(0.25)  & 4.20(0.25)\\
\vmicro              & \kms & 1.43(0.40)  & 1.05(0.45)\\
\vsini              & \kms & 2.4(1.0)    & 2.5(1.1)\\
\vmacro             & \kms & 2.3(1.0)    & 2.6(1.0)\\
${\rm [M/H]}$       & dex  & $-$0.02(0.10)& $-$0.09(0.10)\\
R$_A$/R$_B$         &      & \multicolumn{2}{c}{1.00(0.12)}\\
\multicolumn{4}{c}{{\bf {\sc HERMES} spectra ({\sc SME})\rule{0pt}{11pt}}}\\
\te\rule{0pt}{15pt} & K    & 6\,055(70)  & 6\,069(70)\\       
\logg               & dex  & 4.08(0.12)  & 4.14(0.09)\\       
\vmicro             & \kms & 1.45(0.18)  & 1.48(0.18)\\       
\vsini              & \kms & 2.2(1.8)    & 1.1(3.4)\\        
\vmacro             & \kms & 5.5(0.8)    & 5.3(0.8)\\         
${\rm [M/H]}$       & dex  & 0.03(0.06)  & 0.03(0.06)\\       
R$_A$/R$_B$         &      & \multicolumn{2}{c}{1.08 (fixed)}\\
		\hline\\
		\end{tabular}
		\end{table}

\subsection{Disentangling process }\label{sect:SLD}

Two main alternative approaches are widely used in the community for spectral disentangling, that either operate in the Fourier or wavelength domain. In the former case, two a priori unknown disentangled spectra are optimized along with the orbital elements of the system, minimizing the cost function based on the observed and predicted time-series of the composite spectra (e.g., as implemented in the {\sc KOREL} \citep{1997A&AS..122..581H} and {\sc FDBinary} software packages (fd3) \citep{2004ASPC..318..111I}). The latter approach assumes operations in the wavelength domain where the two separate spectra are typically resolved under the assumption of fixed RVs and flux contribution of the individual binary components (e.g., as implemented in the CRES software package \citep{2004ASPC..318..107I}). A variation of the methods also exists where RVs are optimized along with the spectra of the components, e.g. as implemented in the {\sc Spectangular} software package developed by \citet{2017A&A...597A.125S}.  

We tried {\sc KOREL}, {\sc FDBinary}, and {\sc Spectangular} programs for disentangling the HD\,60803 spectra. Neither of the three delivered the desirable level of accuracy in the disentangled spectra and the final product was found to suffer significantly from the continuum undulations. We attribute the failure to obtain reliable disentangled spectra to the extreme complexity of the input spectra that are represented by the superposition of spectral contributions of two solar-type stars making it extremely difficult to identify spectral line-free continuum points. 
  
Hence, we employ a new technique based on the Spectral Line Deconvolution (SLD) method developed by us. We search for a function $X(v)$ which gives us the observed spectrum $R_{\lambda}$ after convolution with the theoretical profiles of the individual spectral lines $p(v)$: 
\begin{equation}
   R_{\lambda}=\sum a_{\lambda} \cdot (X(v) \ast p(v))\label{eq1}
\end{equation}   

\noindent   
where $a_{\lambda}$ is a set of multiplication factors that account for the fact that the profile of a blend is not necessarily a linear sum of the individual profiles contributing to that blend. Hence, the SLD-function is a normalized broadening profile which includes all global physical and instrumental effects defining the shape and position of a spectral line: radial velocity, rotation, turbulence, spectral resolution. In some sense, SLD is similar to the widely-used Least Squares Deconvolution (LSD) method \citep{1997MNRAS.291..658D}: both methods make use of spectral line masks, yet the definition of the mask is different. The LSD mask is represented by a list of delta-functions (line positions and their central depths) while the method assumes similarity of all considered spectral lines along with their strict linear addition to represent the observed stellar spectrum. The SLD method relies on the mask to contain pre-computed intrinsic profiles for all spectral lines as well as the $a_{\lambda}$-corrections to account for possible non-linear contribution of a given spectral line to the observed blended profile. Eq.~\ref{eq1} can be generalized to spectroscopic binary (SB2) and higher order multiple systems. In the specific case of SB2 systems, we use two line masks and solve for two SLD functions, each one corresponding to the individual component of the binary system. Application to binaries is also associated with the introduction of an extra free parameter, namely flux ratio of the two stars. The flux ratio is wavelength-dependent, however in the particular case of the HD 60803 system the wavelength dependence can be safely ignored given how similar the two components are in terms of their atmospheric and fundamental parameters. 

Disentangled spectra are obtained in an iterative approach, where we (i) compute theoretical line masks for both stars assuming a set of atmospheric parameters (effective temperature \te\, surface gravity \logg\, metallicity, and microturbulent velocity) for each of the components of the binary system, (ii) solve Eq.~\ref{eq1} for SLD functions, optimize for the flux ratio of the two stars, and perform line strength corrections; (iii) perform the convolution of the obtained SLD functions with the improved line masks to obtain the disentangled spectra of both binary components. The obtained disentangled spectra are then analysed with respect to determination of atmospheric parameters of the stars and the whole cycle (i)-(iii) is repeated until no significant changes in the line strengths corrections, the flux ratio, and the atmospheric parameters inferred from the disentangled spectra are recorded. 
We set up the initial values for atmospheric parameters, radial velocities of the binary components and their radii ratio.  Model atmospheres are computed with the {\sc LLmodels} software package \citep{2004AA...428..993S} while the {\sc SynthV} code \citep{1996ASPC..108..198T} is employed in tandem with the atomic and molecular line parameters extracted from the recent version of the {\sc VALD} database \citep{2017ASPC..510..509P} to compute synthetic spectra together with the individual line profiles. 

\begin{figure}           
	\includegraphics[width=0.24\textwidth,clip]{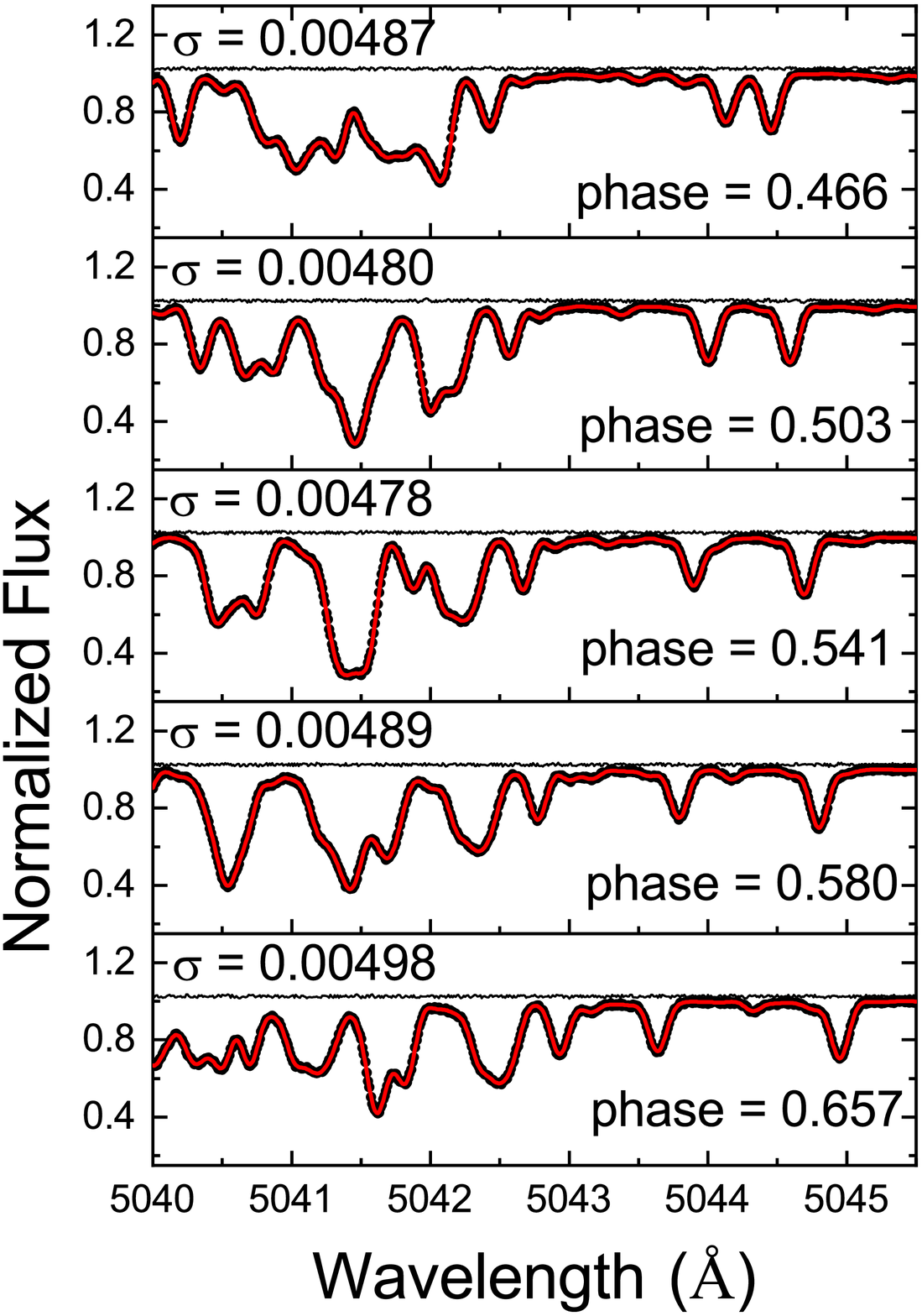}\,\includegraphics[width=0.24\textwidth,clip]{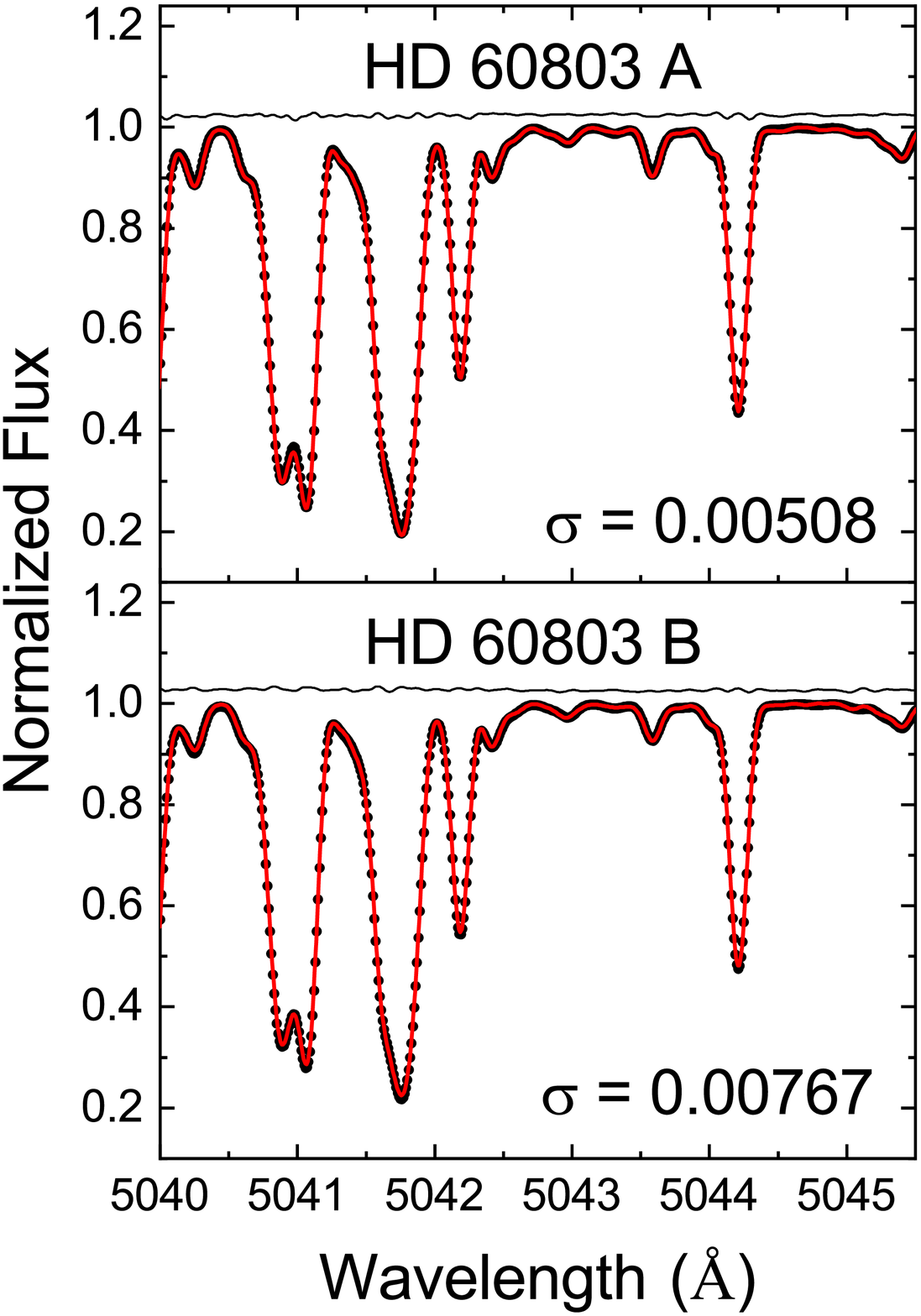}
	\caption{Left columns: Quality of the fit to the test binary spectra of the HD\,60803 system in 5 orbital phases. The composite binary spectra are shown by black dots, the best fit model spectra are shown by red line, the black line refers to the residuals between the composite spectra and the best fit model. \newline
	Right columns: The disentangled test spectra of the components at the same spectral regions (black dots). Red lines shows a fit of the synthetic spectra calculated with the final atmospheric parameters to the individual spectra of the components.} \label{disen-test}
\end{figure}

\begin{figure*}           
	\includegraphics[width=0.48\textwidth,clip]{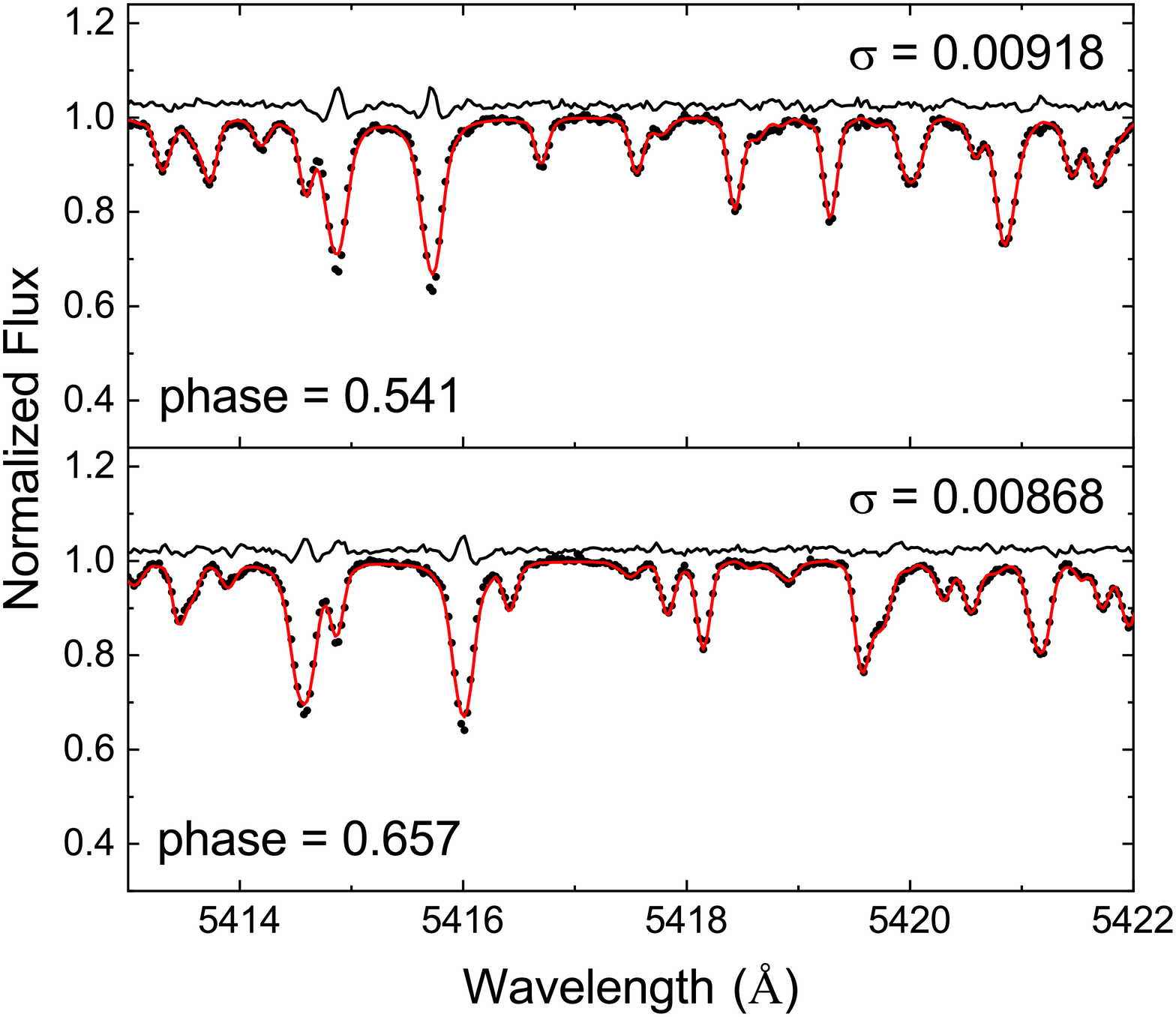}\includegraphics[width=0.48\textwidth, clip]{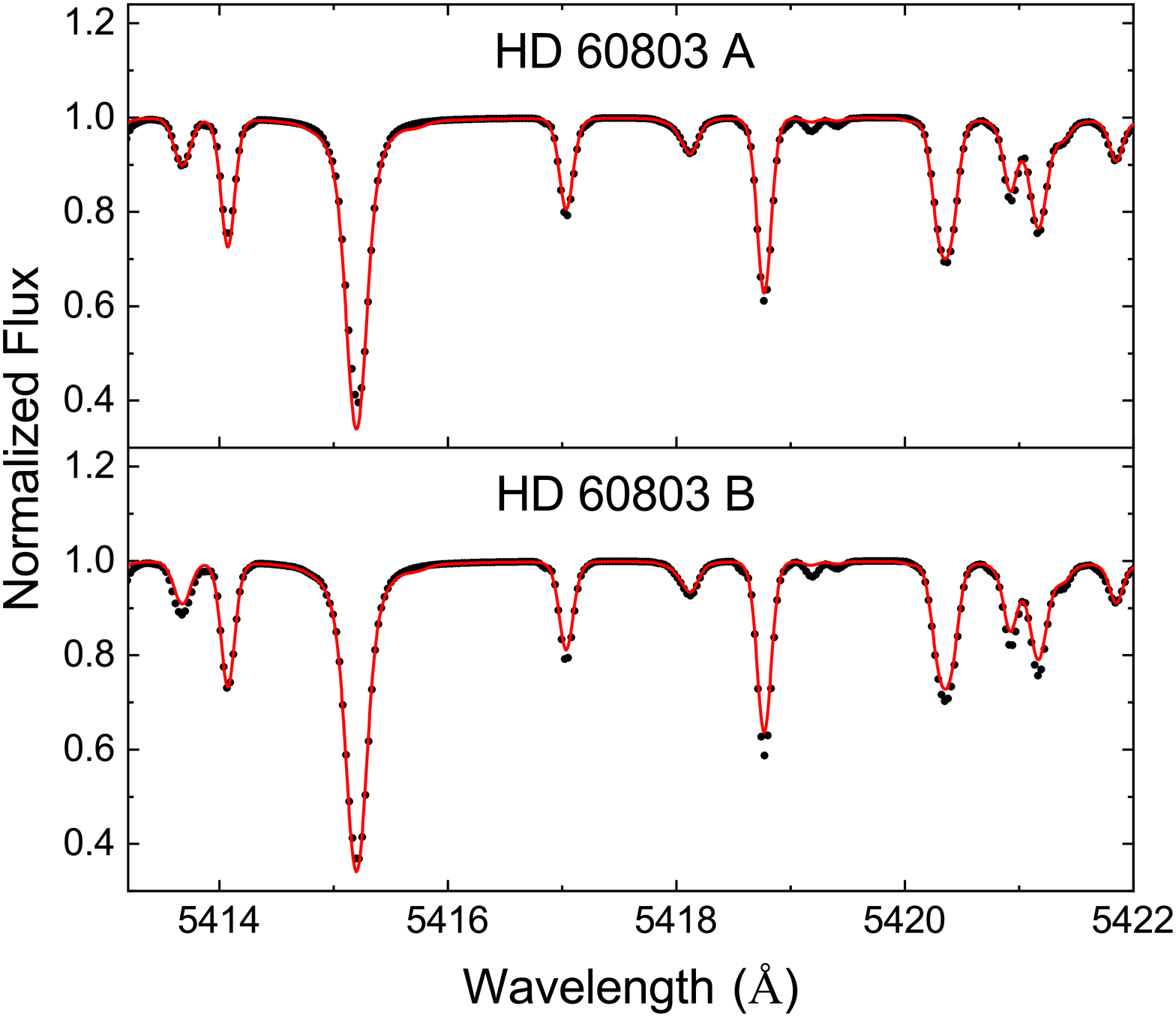}
    \includegraphics[width=0.48\textwidth,clip]{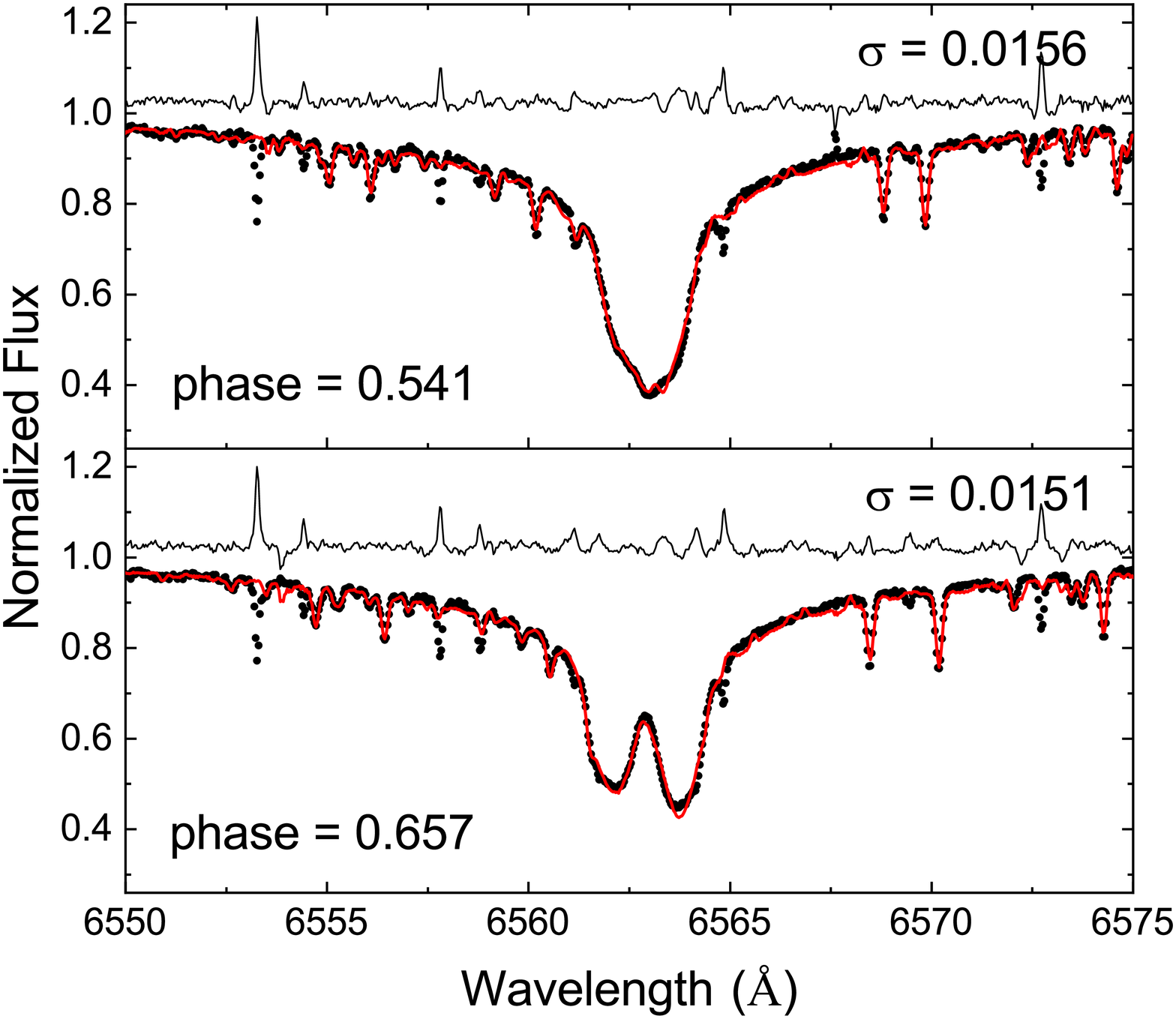}\includegraphics[width=0.48\textwidth, clip]{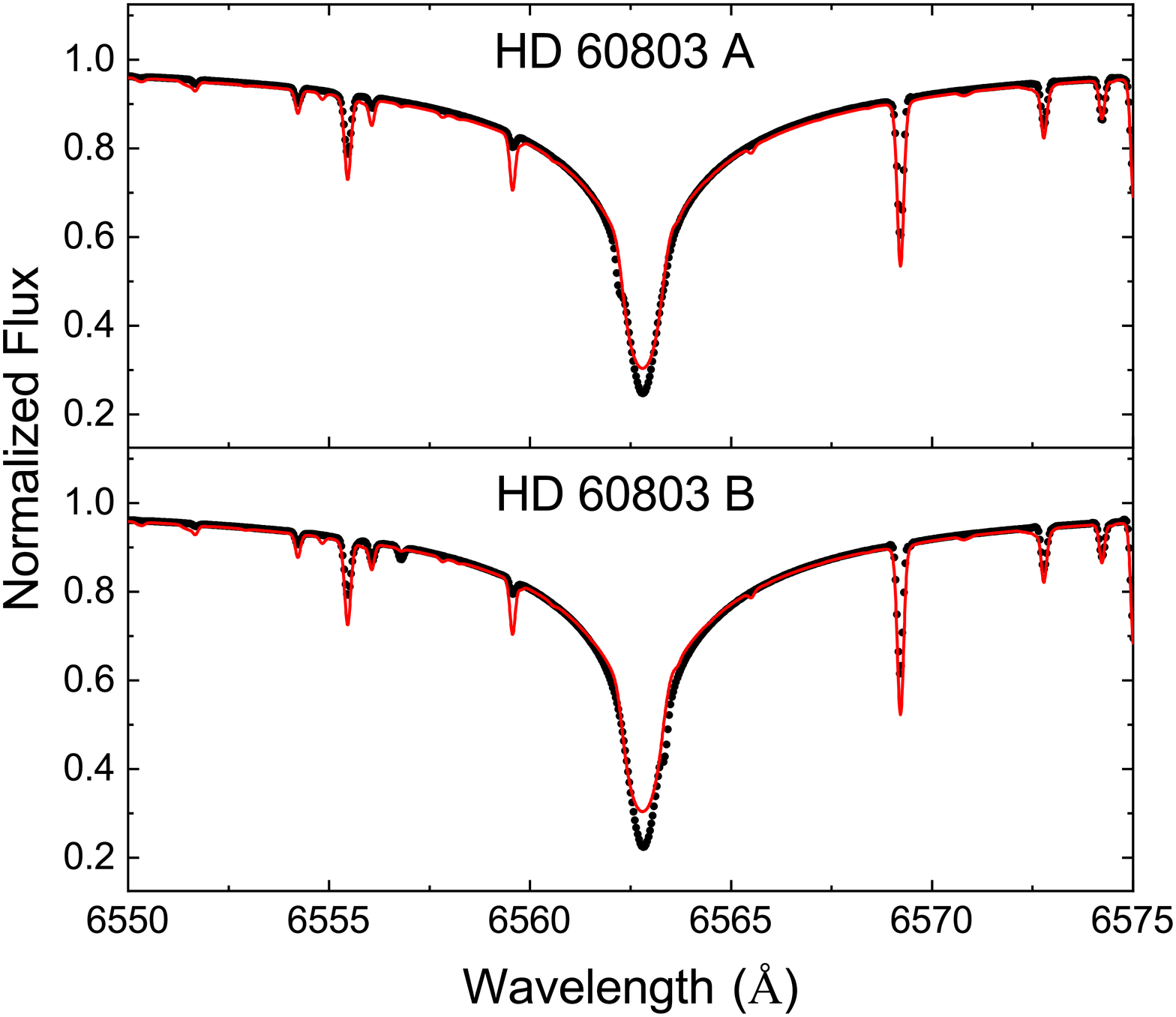}
	\caption{Left columns: Quality of the fit to the observed spectra of the HD\,60803 system in two different orbital phases. Observations are indicated by black dots, the best fit composite profile is shown by red line, the black line refers to the residuals between the observations and the best fit model. The peaks in the residuals on the bottom left panels indicate unsubtracted telluric lines. \newline
	Right columns: The disentangled spectra of the components at the same spectral regions (black dots). Red lines show the fit of the synthetic spectra calculated with the final atmospheric parameters to the individual spectra of the components.} \label{disentangl}
\end{figure*}

The initial parameters were chosen close to those derived from the analysis of {\sc ELODIE} and {\sc HiRes} spectra. These parameters are listed in Table~\ref{tab_SLD} together with the errors in parenthesis, adopted in the disentangling process. For test purposes, we employ artificial binary spectra at different orbital phases. Synthetic spectra of the components expressed in flux units were added together with the shifts in wavelength scale according to the radial velocity of each component. These composite spectra were then convolved with the instrumental profile corresponding to $R$=85\,000 and noise was added to simulate S/N=120. Spectrum disentangling was performed with the atmospheric parameters perturbed within their respective uncertainties. Our tests show that the results of the disentangling process are not sensitive to the variations of atmospheric parameters within the error limits when computing the SLD masks.    
An example of the disentangling of the artificial data set is presented in Fig.~\ref{disen-test}. As for the disentangling of the real observed spectra, one iteration was enough to get the results due to the proper choice of the initial parameters. Fig.~\ref{disentangl} shows the results of the disentangling of the observed spectra of the HD\,60803 system. The SLD-based disentangling is successful in reproducing strong lines with developed wings (e.g. H$\alpha$ line) just as well as weaker or intermediate-strength lines (e.g., Fe\ione~$\lambda$ 5415.2~\AA). Lines broadened by hyperfine splitting are reproduce as well (e.g., Mn\ione~$\lambda$ 5420.3~\AA); an effect of asymmetry caused by isotopic separation is illustrated in Fig.~\ref{fig_Li}. 

\begin{table}
	\centering\tabcolsep 1.5mm\caption{Atmospheric parameters of both stellar components of the HD\,60803 system adopted for testing procedure. Error limits are shown in parentheses.} \label{tab_SLD}
\begin{tabular}{llcc} \hline
	\multicolumn{1}{c}{Parameter\rule{0pt}{9pt}} &
	\multicolumn{1}{c}{Unit} & \multicolumn{1}{c}{Primary(A)} &
	\multicolumn{1}{c}{Secondary(B)}\\\hline
\te\rule{0pt}{15pt} & K    & 5\,900(150) & 6\,000(150)\\
\logg               & dex  & 4.0(0.2)  & 4.1(0.2)\\
\vmicro             & \kms & 2.0(0.2)  & 2.0(0.2)\\
\vsini              & \kms & 3.0       & 3.5   \\
\vmacro             & \kms & \multicolumn{2}{c} {3.0(fixed)}\\
${\rm [M/H]}$       & dex  & $-$0.2(0.1)& $-$0.2(0.1)\\
R$_A$/R$_B$         &      & \multicolumn{2}{c}{1.08(0.05)}\\
		\hline\\
		\end{tabular}
		\end{table}

Spectrum disentangling of HD\,60803 was performed using 5 {\sc HERMES} spectra. 

\subsection{Radial velocity measurements}\label{sect:RV}  

Radial velocities of the components were derived using 2D cross-correlation technique similar to that employed in {\sc TODCOR} code \citep{1993AAS...183.8601M}. This method was applied to all spectra of HD\,60803. A grid of composite spectra were calculated for each orbital phase based on the sums of the individual spectra of the components shifted by a set of corresponding RVs around an expected RV from the orbital solution. Then these synthetic spectra were compared to the observed ones varying RV of each component, and the final RV-values were derived based on the  $\chi^2$ statistics similar to the procedure described above. These values together with the errors are given in Table~\ref{tab1}.
  
\begin{figure}           
	\includegraphics[width=0.4\textwidth]{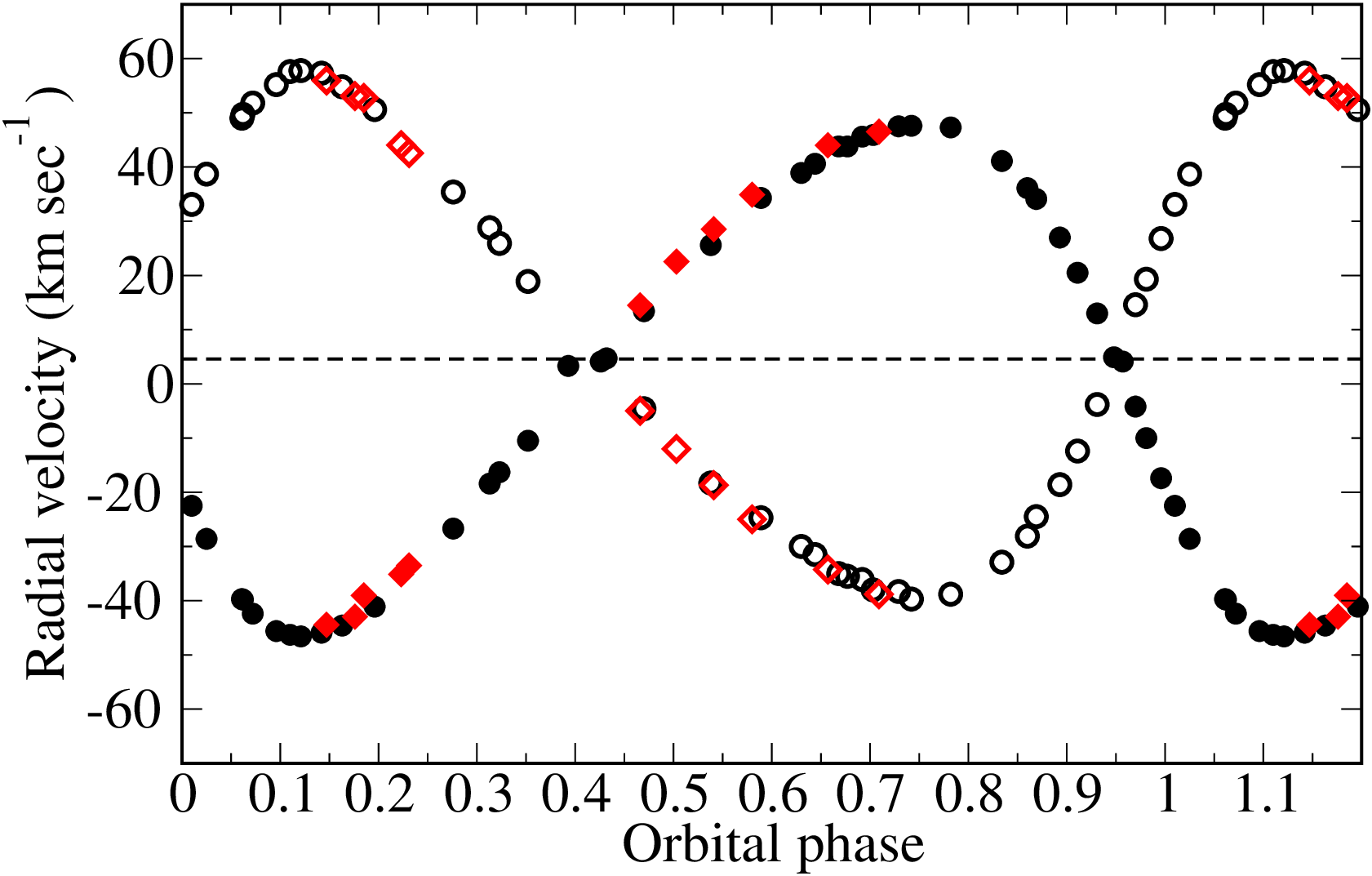}
	\caption{The observed radial velocities of HD\,60803 as a function of the orbital phase. RV values of the primary (HD~60803~A) are shown by the filled symbols, while the open symbols represent RV's of the secondary (HD~60803~B). Literature data from \citet{1997Obs...117..208G} are shown by black circles, results of the current work are shown by red diamonds.} \label{RV}
\end{figure}

Radial velocities from \citet{1997Obs...117..208G} together with the current data are plotted on Fig.~\ref{RV}. RV measurements separated by 22 years are nicely represented by the orbit solution from \citet{1997Obs...117..208G} that provides an evidence for the orbit stability. 

\subsection{Atmospheric parameters and abundance analysis }\label{sect:abn}

The disentangled spectra of both components were processed with Spectroscopy Made Easy ({\sc SME}) spectral package  \citep{1996AAS..118..595V,2017A&A...597A..16P}. This package is designed for automatic determination of the model atmosphere parameters (\te, \logg),  velocity fields (rotation \vsini, micro-\vmicro\ and macroturbulence \vmacro) and for abundance analysis based on the fitting of the theoretical spectra to the observed ones. {\sc SME} employs the careful selection of the spectral features with different sensitivity to the atmospheric parameters (line mask). It works with different grids of model atmospheres: {\sc MARCS} models \citep{2008AA...486..951G}, {\sc LLmodels} \citep{2004AA...428..993S}, and {\sc ATLAS9} models \citep{2003IAUS..210P.A20C}.

SME package  was successfully applied to the study of 1617 FGK-type stars in Planet-search group \citep{2016ApJS..225...32B}. \citet{2016MNRAS.456.1221R} carried out a comparative analysis of the results obtained with {\sc SME} and with other spectroscopic tools employing different model atmosphere grids.
The authors found that effective temperatures derived with {\sc LLmodels} and {\sc MARCS} models agree within $\pm$15--20~K which is less than the uncertainties of temperature determinations. 
It was concluded that {\sc SME} provides reliable atmospheric parameters and element abundances with the reasonable error estimates. 
We used {\sc MARCS} models in our analysis because departures from the local thermodynamic equilibrium (NLTE) for few elements are calculated for this particular grid \citep{2017ASPC..510..509P}.  
   
\begin{figure}           
	\includegraphics[width=0.48\textwidth]{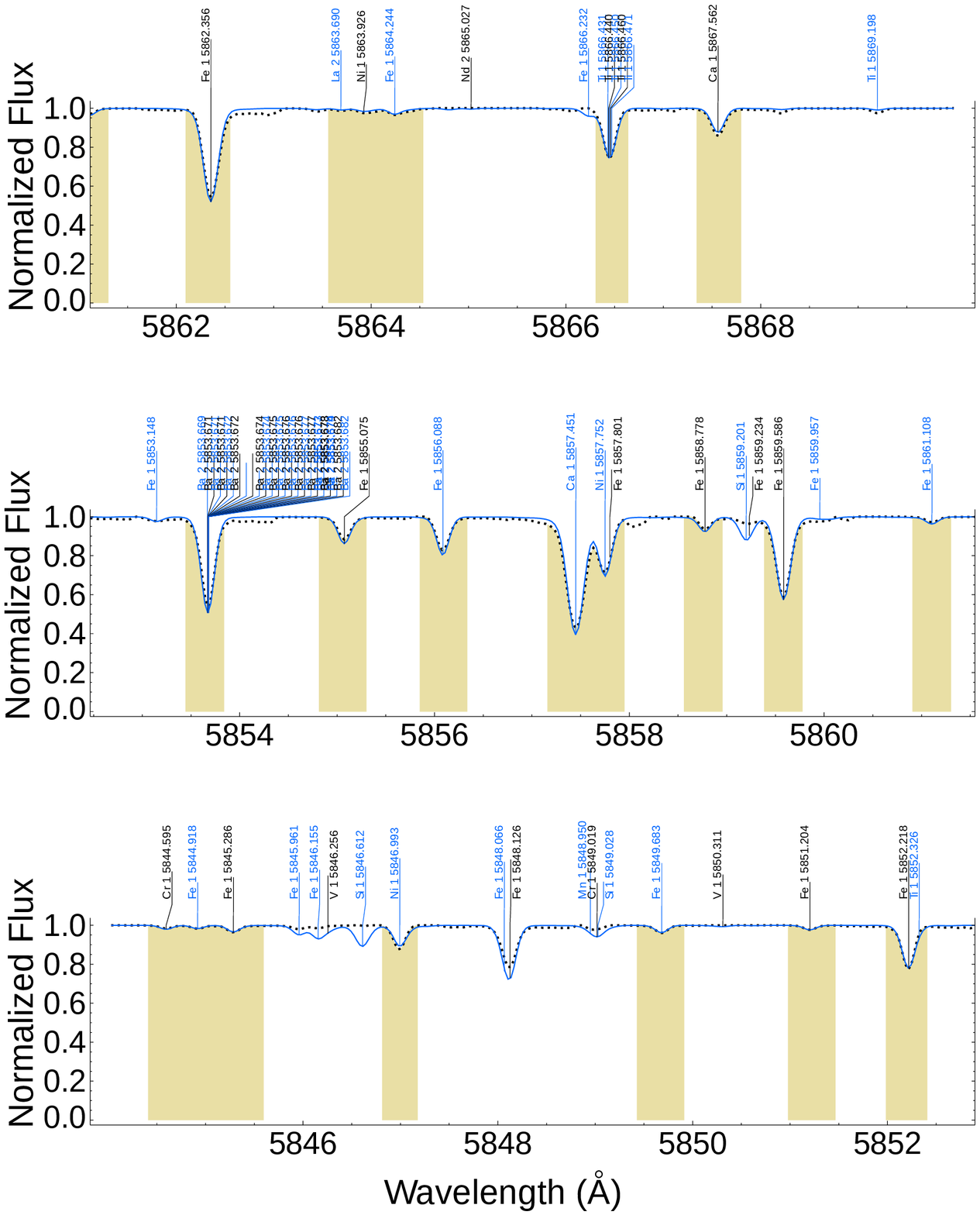}
	\caption{Example of the line mask in 5844-5870~\AA\ spectral region for {\sc SME} fitting procedure. Spectrum of HD\,60803~A is shown by black dots. Line mask is shown by yellow colour. } \label{mask}
\end{figure}

As in paper by \citet{2016MNRAS.456.1221R} we used the spectral regions 4485--4590~\AA, 5100--5200~\AA, 5600--5700~\AA, 6100--6200~\AA. and 6520--6580~\AA\ (H$\alpha$) for fitting procedure. We also implemented four additional regions: 5500-5550~\AA\, 5700-5720~\AA\ (Mg\ione),  5844-5870~\AA\ (Ca\ione, Ba\ii), and 7700-7810~\AA\ (O\ione\ triplet). The same line masks as in \citet{2016MNRAS.456.1221R} were used for the first 5 regions, while we have created new masks for 4 new regions. An example of the mask in 5844-5870~\AA\ region is given in Fig.~\ref{mask}. For all elements but neutral sulphur atomic line parameters were extracted from the recent version of {\sc VALD3} which provides the isotopic and hyperfine structure components of the spectral lines for few elements \citep{2019ARep...63.1010P}. For S\ione\ oscillator strengths were taken from \citet{7952TP}. NLTE was taken into account for O, Na, Ca, and Ba \citep[see][and references therein]{2017ASPC..510..509P}. Atmospheric parameters deduced from the analysis of the {\sc HiRes} spectrum were used as the starting ones in {\sc SME} procedure. We performed 2 steps in SME: in the first step the best-fitting solution was searched for effective temperature \te, surface gravity \logg,
metallicity [M/H], micro- \vmicro\ and macroturbulent \vmacro\
velocities, and projected rotational velocity \vsini. In the second step we fixed parameters derived in the first step and searched the best-fitting solution for abundances of 18 elements from C to Ba. Final parameters are presented in Table~\ref{tab_param}, while element abundances with the uncertainty estimates are given in Table~\ref{tab_abund}. 
Table~\ref{tab_abund} also contains photospheric solar abundances \citep{Asplund,2015A&A...573A..25S,2015A&A...573A..26S,2015A&A...573A..27G} and condensation temperatures taken from \citet[][Table 8]{2003ApJ...591.1220L}. Elemental abundances are given in logarithmic scale relative to the total number of atoms $\log(N_{el}/N_{tot})$.       

Error estimates for all varied parameters and abundances were derived as described in the papers by  \citet{2016MNRAS.456.1221R} and \citet{2017A&A...597A..16P}.  It is based on the analysis of the cumulative distribution for each parameter. Central parts of the corresponding probability distributions are not too far from the Gaussians (see Fig.1 and Fig.2 in \citet{2016MNRAS.456.1221R}). Carbon abundance in {\sc SME} is estimated mainly by fitting numerous weak molecular lines of C$_2$ (Swan system), which are strongly temperature dependent in cool stars. In addition, we derived carbon abundance using five the least blended atomic lines of
C\ione~$\lambda\lambda$~5052.1, 5380.3, 7111.5, 7117.0, 7119.7~\AA\ which were corrected for NLTE effects as described by \citet{2016MNRAS.462.1123A}. Individual line abundances were calculated with a NLTE spectrum fitting software described in \citet{2019ASPC..518..247T}. 
Abundances of the elements S and Zn which lines did not fall into spectral intervals chosen for {\sc SME} were calculated with the same software. We used S\ione~$\lambda\lambda$~6052.6, 6757.2~\AA\ and Zn\ione~$\lambda\lambda$~4722.1, 4810.5~\AA\ spectral lines for S and Zn, correspondingly. Although we use slightly different synthetic spectrum codes in {\sc SME} and for individual line abundance calculations, they give abundance results which agree within 0.02~dex or better \citep{2014dapb.book..277R}. For example, mean carbon LTE abundance based on {\sc SME} calculations results in $(\log(N_{C}/N_{tot})_A=-3.61\pm0.05$ against $-3.62\pm0.05$ derived from individual line fits.  It is worth to mention that the final parameter and abundance uncertainty estimates in {\sc SME} include both systematic uncertainty (model limitations) and the observational one, while standard deviations are given for C, S and Zn.

Rather strong observed features at the position of Li\ione~$\lambda$6707.8~\AA\ line allows us to derive accurately Li abundance in both components taking into account the isotopic \citep{REB} and hyperfine \citep{BBE,OAO} splittings. Li abundance was derived with {\sc SME} version which includes Li$_{NLTE}$ grids calculated according to \citet{2009A&A...503..541L}.  
NLTE corrections are small, not exceeding -0.04~dex. NLTE Li atmospheric abundances are given in Table~\ref{tab_abund}. The uncertainty refers to the temperature variations. We derived $\log\epsilon_{Li}$=2.94 (HD~60803~A) and 2.89 (HD~60803~B),
that is much higher than the solar photospheric Li abundance $\log\epsilon_{Li}$=1.05 \citep{Asplund} and lower than the meteorite abundance $\log\epsilon_{Li}$=3.28 \citep{2003ApJ...591.1220L}.  Li abundance in HD\,60803 agrees within the error bars with  $\log\epsilon_{Li}$=3.09$\pm$0.13 derived for young F-G stars in the same effective temperature domain \citep{2011MNRAS.410.2526B}, and with the Li abundance in field stars of low/unknown chromospheric activity lying in temperature range of 6000-6300~K \citep[see Fig.2 in][]{2000MNRAS.316L..35R}. It exceeds the primordial Li abundance $\log\epsilon_{Li}=2.72\pm0.06$ \citep{2012ApJ...744..158C}.  

\begin{figure}           
	\includegraphics[width=0.48\textwidth]{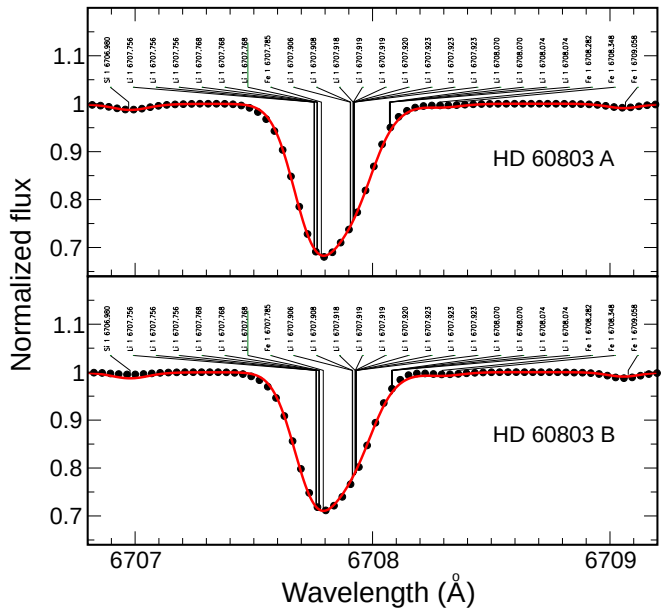}
	\caption{Comparison between the observed spectra of the HD\,60803 system components in Li\ione~$\lambda$6707.8~\AA\ line region (black dots) and the synthetic spectra (red lines).} \label{fig_Li}
\end{figure}

\begin{table}
\tabcolsep=1.0mm
	\caption{Elemental abundances ($\log(N_{el}/N_{tot})$) in the atmospheres of the HD\,60803 components and their differences. The uncertainty in the last digits is given in parentheses. Photospheric solar abundances taken from \citet{Asplund,2015A&A...573A..25S,2015A&A...573A..26S,2015A&A...573A..27G} are displayed in the 5th column.  The last column presents element condensation temperature (\CT) \citep{2003ApJ...591.1220L}.}	\label{tab_abund}
	\small
	\centering
	\begin{tabular}{l c|c|r|r|r}
		\hline \hline Element & \multicolumn{4}{c|}{Abundance}& \CT (K)  \\ 
		\hline & HD\,60803~A & HD\,60803~B & A-B & Sun\\
		\hline 
	    Li  &{\bf -9.10(05)} &{\bf -9.15(05)} & {\bf 0.05} &-10.99 &1135  \\      
        C(atom)&{\bf -3.64(05)} &{\bf -3.63(04)} & -0.01 & -3.61 &  40  \\ 
		C(mol) & -3.59(11) & -3.59(12) &  0.00 & -3.61 &  40  \\ 
		O 	    & -3.39(06) & -3.39(07) &  0.00 & -3.35 & 180  \\  
		Na 	   & -5.78(03) & -5.74(04) & -0.04 & -5.83 & 958  \\  
		Mg	    & -4.43(03) & -4.42(03) & -0.01 & -4.45 &1336  \\ 
		Si     & -4.50(08) & -4.50(08) &  0.00 & -4.53 &1310  \\  
		S 	    & -4.93(01) & -4.90(02) & -0.03 & -4.92 & 664  \\   
		Ca 	   & -5.63(07) & -5.64(07) &  0.01 & -5.72 &1517  \\ 
		Sc 	   & -8.80(08) & -8.79(08) & -0.01 & -8.88 &1659  \\ 
		Ti 	   & -7.03(06) & -7.04(06) &  0.01 & -7.11 &1582  \\ 
		V 	    & -8.01(04) & -8.01(04) &  0.00 & -8.15 &1429  \\   
		Cr 	   & -6.36(08) & -6.35(07) & -0.01 & -6.42 &1296  \\   
		Mn 	   & -6.58(05) & -6.58(08) &  0.00 & -6.62 &1158  \\   
		Fe	    & -4.52(08) & -4.52(08) &  0.00 & -4.57 &1334  \\   
		Co 	   & -7.03(10) & -7.06(09) &  0.03 & -7.11 &1352  \\   
		Ni     & -5.76(08) & -5.78(08) &  0.02 & -5.84 &1353  \\  
		Cu     & -7.87(08) & -7.87(07) &  0.00 & -7.86 &1037  \\ 
		Zn     & -7.45(02) & -7.48(05) &  0.03 & -7.48 & 726  \\  
		Y 	    & -9.91(09) & -9.89(10) & -0.02 & -9.83 &1659  \\  
		Zr 	   & -9.45(02) & -9.43(02) & -0.02 & -9.45 &1741  \\   
		Ba     & -9.80(02) & -9.80(03) &  0.00 & -9.79 &1455  \\ 
		\hline 
 \end{tabular}		 
\end{table}	

\begin{figure*}
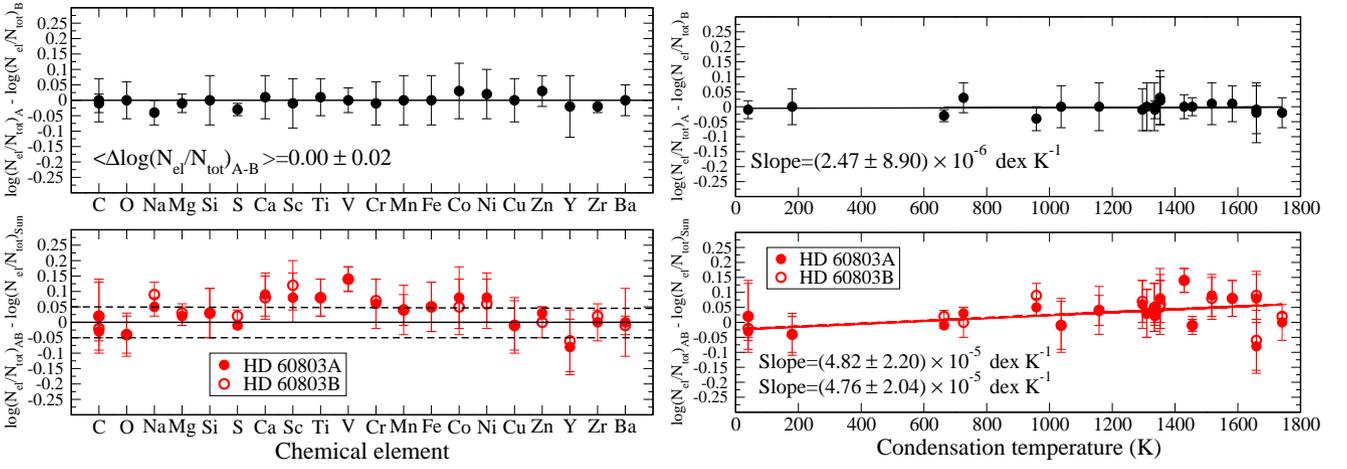
           
	\includegraphics[width=0.48\textwidth]{Abun_diff.eps}\hspace{2mm}\includegraphics[width=0.48\textwidth]{Abun_dif_CT.eps}
	\caption{Abundance difference between the components A-B (upper panels) and relative to the solar abundances of the components of HD 60803 (lower panels) in dependence on the chemical element (left panels) and condensation temperature \CT\ (right panels). A typical range of the uncertainty $\pm$0.05~dex in solar abundance determinations is indicated by dashed lines (left bottom panel). Linear regressions in abundance versus \CT\ are shown by solid and dashed lines on the right bottom panels.} \label{FigN}
\end{figure*}

Element abundances (without Li) in the atmospheres of the HD\,60803 components relative to photospheric solar abundances are displayed in Fig.~\ref{FigN} (bottom panels), where they are shown in dependence on the atomic number (element name) and on the element condensation temperature \CT. The upper panels demonstrate the same correlations but for the abundance difference between the components. Our results show that there is no significant deviations of the component abundances from the solar ones as well as an absence of any significant correlation with the \CT. The formal abundance difference between components $\Delta\log{\rm (N_{el}/N_{tot})_{A-B}}=0.00\pm0.02$. The formal slopes in abundance versus \CT\ for both components are $(4.82\pm2.20)\times10^{-5}\,\rm {dex\,K^{-1}}$ (pimary) and $(4.76\pm2.04)\times10^{-5}\,\rm {dex\,K^{-1}}$ (secondary). It lies between the slopes of the similar correlations found by \citet{2009ApJ...704L..66M} for the solar analogues with/without giant planets relative to the Sun.

\section{Evolutionary status of HD\,60803 system}\label{sect:evol}

\begin{figure*}
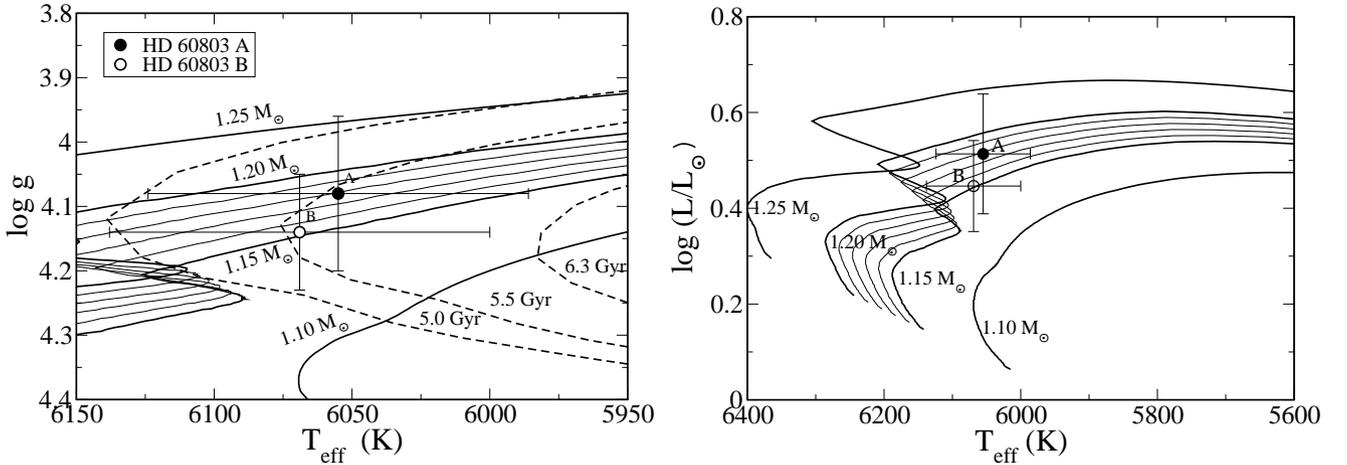
           
	\includegraphics[width=0.48\textwidth]{HD60803_evol_age_2012_BW.eps}\hspace{2mm}\includegraphics[width=0.48\textwidth]{HR_diagram.eps}
	\caption{Left panel: Position of the HD\,60803 components on the (\te\--\logg\ diagram. Evolutionary tracks in step of 0.05\Msun\ taken directly from the non-rotating, solar metallicity grid \citep{2012A&A...537A.146E} are shown by thick solid lines; those in 1.15$\leq$\Msun$\leq$1.20 mass range are interpolated with step 0.01\Msun\ and shown by thin solid lines. Three isochrones for $t$=5.0, 5.5, 6.3~Gyr are shown by dashed lines.\newline
	Right panel: Position of the HD\,60803 components on the HR-diagram.} \label{fig:isoc}
\end{figure*}

Based on the derived atmospheric parameters \te\ and \logg\ we displayed both components of HD\,60803 system on the evolutionary tracks and isochrones calculated by \citet{2012A&A...537A.146E} for non-rotating solar metallicity models (Fig.~\ref{fig:isoc} -- left panel). 

We estimated masses of the components as $M_A=(1.18\pm0.08$)\Msun, $M_B=(1.15\pm0.06$)\Msun, which gives a mass ratio $q=M_A/M_B$=1.025$\pm$0.10 in agreement with the mass ratio $q=1.019\pm0.003$ derived from the orbit solution \citep{1997Obs...117..208G}.The age of the system is estimated as $t=5.5\pm0.5$~Gyr.  Similar results  were obtained with the evolutionary tracks and isochrones from single-star MESA (Modules for Experiments in Stellar Astrophysics) theoretical stellar evolution model grid\footnote{\url{http://waps.cfa.harvard.edu/MIST/}} for solar abundance and zero rotational velocity  \citep{2016ApJ...823..102C}: $M_A=1.13$\Msun\ $M_B=1.11$\Msun, $t=5.8$~Gyr. 
Derived surface gravities and estimated masses allow us to calculate radii of both components as
$R_A=(1.64\pm0.23$)\Rsun, $R_B=(1.51\pm0.16$)\Rsun. Their ratio $R_A/R_B=1.09\pm0.19$ is fully consistent with the ratio 1.08 found in disentangling procedure. The luminosities are $\log(L/L_{\odot})_A=0.51\pm0.13$ and $\log(L/L_{\odot})_B=0.45\pm0.10$, respectively. The position of the components of the HD~60803 binary system on HR-diagram is shown in Fig.~\ref{fig:isoc} (right panel). While both stars terminated core hydrogen burning phase, they did not evolve to the phase of first deep mixing, so their surface chemistry did not change from the original one.
Mass estimates together with the orbit solution for $m_{A,B}sin^{3}i$ allow us to estimate orbital inclination as $i=78^\circ\pm4^\circ$ \citep[][evolutionary tracks]{2012A&A...537A.146E} or $i=83^\circ\pm4^\circ$ \citep[][evolutionary tracks]{2016ApJ...823..102C}. According to \citet{1997Obs...117..208G} the orbital inclination should be within $3^\circ$ of 90$^\circ$ for an eclipse to take place, therefore our results give a minor probability for an eclipse. 

An estimated age of the HD\,60803 system $t=5.5\pm0.5$~Gyr corresponds to the age of the oldest open clusters where Li abundances was derived for different ranges of effective temperature \citep{2005A&A...442..615S}. Both components of HD\,60803 system fall into hotter temperature domain 6200$\pm$150~K where $\log\epsilon_{Li}$=2.65--2.55 in clusters of $t\approx$2--5~Gyr. 

\section{Conclusions}\label{sect:Conclusions}
Employing different methods we performed detailed spectroscopic analysis of the SB2 system HD\,60803 based on high resolution spectra. New procedure SLD was shortly introduced and successfully applied to disentangle spectra of a binary system. Two different methods: {\sc gssp\_composite} module of the {\sc gssp} program package and SLD disentangling procedure together with the {\sc  SME} package were applied to analysis of spectral observations obtained with different spectrographs. The first method was applied to a single {\sc ELODIE} and {\sc HiRes} spectra while the second method was used to disentangle a set of binary spectra observed at different orbital phases and 
to analyse individual spectra of binary components. Model atmosphere parameters of the binary components derived independently by both methods agree very well that gives a credit to further abundance analysis based on individual component's spectra.    
According to our analysis both components of the binary system HD\,60803 are slightly evolved solar composition late F-type stars with the masses $\sim$1.15\Msun. They have the same atmospheric abundances that was expected for the stars formed from one cloud. Their relative-to-solar abundances correlate with \CT\, the same way as was found by \citet{2009ApJ...704L..66M} for the solar analogs with/without detected giant planets.  

\section*{Acknowledgments}
The authors acknowledge the support of Ministry of Science and Higher Education of the Russian Federation under the grant 075-15-2020-780 (N13.1902.21.0039). 
Part of the research has received funding from the KU~Leuven Research Council (grant C16/18/005: PARADISE), the Research Foundation Flanders (FWO) under grant agreement G0H5416N (ERC Runner Up Project), as well as the BELgian federal Science Policy Office (BELSPO) through PRODEX grant PLATO.

\section*{Data Availability Statements}
The data underlying this article will be shared on reasonable request to the corresponding author.

\bibliography{reference,references_tsymbal}
\bibliographystyle{mnras}
\label{lastpage}
\end{document}